\documentstyle[epsfig]{mn}
\def\parhang{\noindent\hangindent=0.4 true in \hangafter=1}
\newcount\notenumber
\def\clearnotenumber{\notenumber=0}
\def\note{\advance\notenumber by1 \footnote{$^{\the\notenumber}$}}
\clearnotenumber
\bigskip

\title{\bf The Small-scale Structure in Interstellar HI: A Resolvable Puzzle}

\author[Avinash Deshpande] 
{Avinash A. Deshpande$^{1,2}$\\ 
$^1$ Raman Research Institute, Bangalore 560 080 INDIA : desh@rri.ernet.in \\
$^2$ Physics Department, New Mexico Institute of Mining \& Technology, Socorro, NM 87801 USA }
\begin{document}
\date{}
\maketitle

\begin{abstract}
During the past decade or so, measurements of Galactic HI 
absorption using VLBI against extra-galactic sources, as well as multi-epoch 
observatios in pulsar directions, have detected small-scale transverse 
variations corresponding to tens of AU at the distance of the absorbing 
matter. Hitherto these measurements have been interpreted as small-scale 
structure in the HI distribution with densities $n_{HI}\sim 10^4-10^5$ 
cm$^{-3}$, orders of magnitude greater than those of the parsec-scale 
structure. Naturally it is difficult to imagine how such structures could 
exist in equilibrium with other components of the ISM.

In this paper we show that structure on all scales contributes to the 
differences on neighbouring lines of sight, and that the observed 
differences can be accounted for by a natural extension 
of the distribution of irregularities in the distribution of HI opacities 
at larger scales, using a single power law.
This, in our opinion, should {\it put an end to the decades long puzzle of the so-called
small-scale structure} in HI and other species in the Galaxy.
\end{abstract}


\pagestyle{myheadings}
\markboth{Avinash Deshpande}{The Small-scale Structure in HI: A Resolvable Puzzle}

\section{Introduction}

 The warm component of the Galactic neutral atomic hydrogen (HI), studied extensively through
its 21-cm line emission using single-dish measurements, shows largely uniform
distribution and has revealed the large-scale
structure of our galaxy. The cold atomic component has also been probed using
the 21-cm absorption observable in the spectra of {\it bright} continuum sources in the background.
The earliest interferometric study towards the bright supernova remnant Cas-A 
(Clark, Radhakrishnan \& Wilson 1962; Clark 1965) revealed the presence of structure finer than
known earlier in cold HI. Later, many aperture-synthesis observations of the Perseus-arm
features detected structures down to the resolution limit (an arc-minute) in these observations
(Greisen 1973; Bregman et al. 1983; Schwarz et al. 1986). These and various other indications had
suggested that the HI gas in our galaxy is organized on, and maintains, a hierarchy of scales
from 1 kpc to at least 1 pc. Contribution from scales much smaller than the parsec-scale
was expected to be a tiny fraction of the total (see, for example, Dickey \& Lockman 1990). 
There were no serious difficulties in understanding this picture.

 However, subsequent HI absorption studies  using VLBI observations of extra-galactic sources
(as background sources) triggered what has
remained as a puzzle for a few decades. Dieter, Welch \& Romney (1976) were the first to
note variations in the HI opacity on a scale smaller than 0.16'' (in the direction of 
3C147).  This they interpreted as implying a 
cold-HI cloud size smaller than 70 AU and a volume density
in the cloud of  $\sim 10^5$ $cm^{-3}$. More than a decade later, Diamond et al.
(1989) reported more VLBI observations on 3C147 as well as two other extra-galactic sources,
supporting the conclusions of Dieter, Welch \& Romney (1976). They interpreted their findings
as suggesting linear diameters of the absorbing clouds to be as small as $\sim$25 AU and 
correspondingly high densities. More elaborate, recent VLBI observations (Davis, Diamond \& Goss 1996;
Faison et al. 1998; Faison 1999) confirm some of the earlier reports of
opacity variations on small transverse scales, while in some other cases find no detectable
variation across different components of the
background sources.

A new technique employing multi-epoch HI-absorption measurements in pulsar directions was
suggested by Frail et al. (1991) as well as by Deshpande et al. (1992) independently.
Based on an extensive multi-epoch study of seven pulsars (sampling various directions in
the galaxy), Frail et al. (1994) reported optical-depth changes of $\le$0.1 on the small
spatial-scales of 5 AU to 100 AU and concluded that a significant fraction 
(10\%-15\%) of the cold HI gas is in such small-scale structure.
 
 Naturally, there are serious difficulties
about such a structure being in pressure equilibrium with the other components of the medium
given the estimated volume density being so high,
and hence about what processes would generate and help maintain such apparently commonly 
encountered structure. 
In a recent paper, Heiles (1997) has summarized the main results from these observations
of the so-called small-scale structure in the interstellar HI (as well as from optical
observations of interstellar absorption lines of NaI \& CaII). 
The paper also points out that existence of such a tiny-scale atomic structure
would imply, under conventional interpretation, over-abundance of H$_2$ leading to very large 
extinction.
To ease such difficulties, Heiles (1997) has proposed geometric solutions invoking structures 
consisting of cold, dense curved filaments or sheets (that line-up along sight-lines)
to explain the observed variations in the HI-opacity, but with moderate
values of the implied volume densities.
On the other hand, Dickey \& Lockman (1990), based on many arguments, conclude that while small-scale
structure does exist, it is only a tiny fraction of the total HI column density along any
sight-lines. 

The two important inter-related questions raised by the apparently puzzling detections of the opacity
variations over small spatial scales, and often asked, are the following.
a) Does the atomic medium resemble the diffuse ionized component ? That is, does it
have a power-law distribution of sizes ? If so, is the behavior similar over the whole
range of scales probed ?  
b) Is the AU-sized structure only peripherally related to the parsec or larger scale structure
in HI ? That is, does it represent a physically distinct population of structures ?

In this paper we show that the observations have been misinterpreted, and 
that the observed small-scale structure is not at all unexpected. The 
observed opacity differences are consistent with a single power-law 
description of the distribution of HI opacities in the interstellar medium, 
and almost all scales contribute significantly to the observed differences. 
It was incorrect to assume that the observed structure must be due to 
high-density clouds whose longitudinal dimension is the same as the 
separation between the lines of sight at distances comaprable with the 
distances of the absorbing matter. 

\section{What do we actually measure in the VLBI and multi-epoch pulsar observations ?}

\subsection{Effects of the source structure and telescope filter function}

The situation in the HI-opacity-variation measurements using VLBI on extra-galactic sources
and the multi-epoch pulsar observations\footnote{For 
further discussion, we will treat, without loss of any details,
the multi-epoch pulsar measurements as equivalent to those made against a background source
consisting of as many number of incoherent component sources as the number of epochs, and 
each component location defined by the apparent pulsar-direction at the corresponding epoch.
Due to the interstellar scattering, the size of the component sources may appear larger 
compared to that intrinsic to the pulsar radiation.
Although the pulsar observations are made usually with a single-dish, for the above
equivalence to be complete, we consider the angular resolution to match 
the scatter broadened size.}
 can be analyzed by identifying three important
ingredients that dictate what actually we would measure; namely, 1) structure of the background
source, 2) structure in the absorbing gas,  and 3) the spatial
frequency filter function of the telescope.
As for their frequency dependence (over the observing bandwidths that are usually small
compared to the centre frequency),
the first quantity can be assumed to be constant, the third one will vary only slightly but 
predictably, while the second quantity can vary considerably from channel-to-channel in frequency
(or velocity). 

 In the continuum channels of the observed band (trivial case of zero-opacity), 
the source structure apparent to the observer is of course the ``true'' structure of 
the background source.  Whereas in the line channels,
which are the ones of interest, the apparent source structure is modified by the opacity structure.
The apparent structure is always a product of the two ``true'' structures, making it, in general, 
appear finer (and consequently extending its visibility range  to higher spatial-frequencies) 
compared to the individual ones.
Two instructive cases are; namely, 1) Uniform finite opacity and VLBI-scale 
structure in the background source and 2) small-scale structure in opacity and uniform brightness
background source. And, let us view these two situations, say, using one VLBI baseline.
In the first case, even though the opacity is uniform (i.e. has no structure), the absorption will
be {\it visible} as long as the ``true'' source structure (which it mimics) has finite visibility 
at a given baseline. 
Thus, the absorption ``visibility'' on a given baseline does not necessarily suggest a structure
in opacity on the corresponding angular (or the related spatial) scale. 
In the second case, the continuum visibility would be zero, but there may be finite visibility
in the line channels. So, any observed ``visibility'' should be directly attributed to a structure
in opacity. In fact, it provides a reasonably pure measure of the power at the 
corresponding spatial scale
in the spectrum characterizing the distribution of opacity (particularly at small optical
depths\footnote{ Although we have used absorption and opacity as analogous to
each other, what the measurements respond to is the modified source structure. And 
the fractional difference in the apparent
structures in two spectral channels gives the structure in fractional absorption from which
the opacity is to be computed.  In this context, the particular non-linear
correspondence between the depth of absorption and the opacity (optical-depth) should be
borne in mind, particularly at large opacities.}).   
Green (1993) has indeed made such interferometric measurements
of the 21-cm emission line to directly sample the associated power spectrum at 
discrete spatial frequencies and Lazarian (1995) has presented a technique to study
the underlying 3-D characteristics using such measurements.
We will return to this case again later.

In reality the situation is somewhere in between the two relatively simple extremes and hence
needs even more care in interpreting the measurements. An essential step then involves proper
imaging of the apparent structures in continuum- and in line-channels and using the comparison
to estimate opacity in the usual way. The opacity can be estimated only over the extent of
the background source imaged. The opacity estimates in closely spaced directions can be
compared as is done in the studies using pulsar-probes and in more recent VLBI studies 
(Davis, Diamond \& Goss 1996; Faison et al. 1998; Faison 1999) that have carefully imaged
the opacity distributions.

\subsection{Expected differences in the opacity \& contributing scales}

 Now, we ask and try to answer two crucial questions,
{\it a) What is the magnitude of opacity differences that we would expect to observe between
 a given pair of  sight-lines ?} and 
{\it b) what scale(s) from the opacity distribution should be considered as contributing to the 
observed opacity differences between two given sight-lines ?} 

Let $\tau_v(x,y)$ represent a two-dimensional distribution of opacity in the transverse coordinates
(x,y) for a given velocity channel. 
For simplicity, let us choose the transverse spatial separation corresponding to the angular separation
between a given pair of {\it thin} sight-lines at the HI-screen distance to be along the x-axis 
and denoted by
$x_o$. The rms value of the opacity difference ($\Delta\tau_v(x_o)$) expected to be 
observed is given simply
by the square-root of the structure function of $\tau_v(x,y)$ at a spatial scale of $x_o$.
Analytically, this can be expressed as\\
     $(\Delta\tau_v(x_o))^2$ = $S_{\tau_v}(x_o)$ = $< (\tau_v(x,y) - \tau_v(x-x_o,y))^2 >$\\
where $<>$ denotes ensemble average of the quantity over all $(x,y)$.

To examine the contributing scales to this opacity difference at separation $x_o$, we consider the
power spectrum ($P_{\tau}(f_x,f_y)$) as a function of the spatial frequencies ($f_x,f_y$) 
(corresponding to spatial scales $1/f_x,1/f_y$)
associated with the distribution $\tau_v(x,y)$, such that $P_{\tau}$ is the Fourier transform
of the autocorrelation function of $\tau_v(x,y)$. If the average power spectrum can be
described as  a power-law, i.e. $<P_{\tau}(f_s)> = P^o_{\tau_v} f^{-\alpha}_s $ (where 
$f_s = \sqrt{f^2_x + f^2_y}$ and such that $\alpha$ is positive) then for $2<\alpha<4$, 
the structure function 
can also be described as a power law such that $S_{\tau_v}(x_o) = S^o_{\tau_v} {x_o}^{\alpha - 2}$
(Lee \& Jokippii 1975). 
Then from the above equation, it follows that
$\Delta\tau_v(x_o) = \Delta\tau^o_{v} {x_o}^{\frac{\alpha-2}{2}}$. 

%Fig:1,2,3

\begin{figure}
\epsfig{file=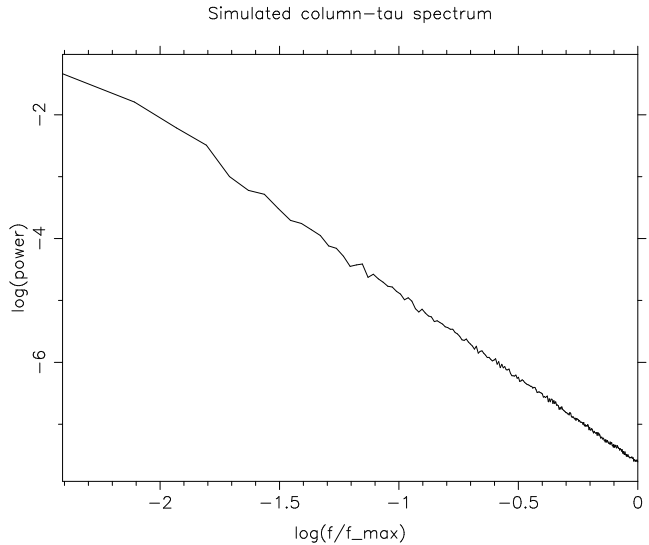,height=8.5cm,angle=-90}
\caption{
An azimuthally averaged power spectral version of a simulated  2-d spectrum
of opacity distribution with $\alpha = 2.75$ 
as suggested by the data in the Cas-A direction (DDG 2000).
The y-axis is calibrated to indicate the power contributed by each
spatial frequency to the $\tau$ distribution.
\label{fig:fig1}}
\end{figure}

\begin{figure}
\epsfig{file=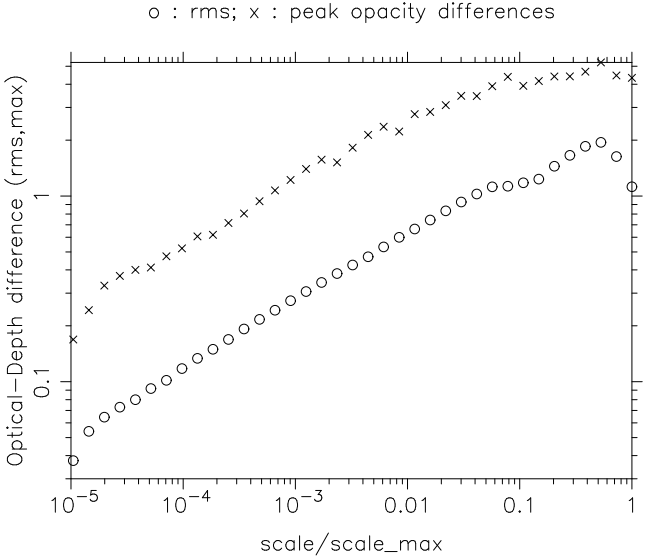,height=8.5cm,angle=-90}
\caption{Two quantities (related to the magnitude of differences expected
as a function of the transverse scale probed)
estimated from 1-d simulations of optical-depth
($\tau$) distribution are shown. The trends with symbols o,x indicate
respectively the rms (i.e. square-root of the structure function) 
and the typical peak difference magnitudes. 
The 1-d distribution of opacity over 5 orders of magnitude in scales (1:$2^{18}$)
was obtained  from a simulated spectrum in 1-d with a power-law index of 1.75
(corresponding to a 2-d equivalent index of 2.75). Due to the computational
requirements being too high, the 2-d case was not attempted on the same scale-range.
However, the 2-d simulation referred to in
Fig. 1 were used to confirm consistency of the results over a range
of longer scales spanning  two orders of magnitude.
The Y-axis values here are calibrated such that the maximum scale may be equated to 4 pc 
(see text for details). The $\tau$ distribution used was
obtained by Fourier transforming a simulated spectrum in the spatial frequency domain;
The repetitive nature implicit in the Fourier transforms can affect
estimation on scales close to the transform length. Hence, the above estimates are made
only for scales that are less than half of the transform length.
\label{fig:fig2}}
\end{figure}

\begin{figure}
\epsfig{file=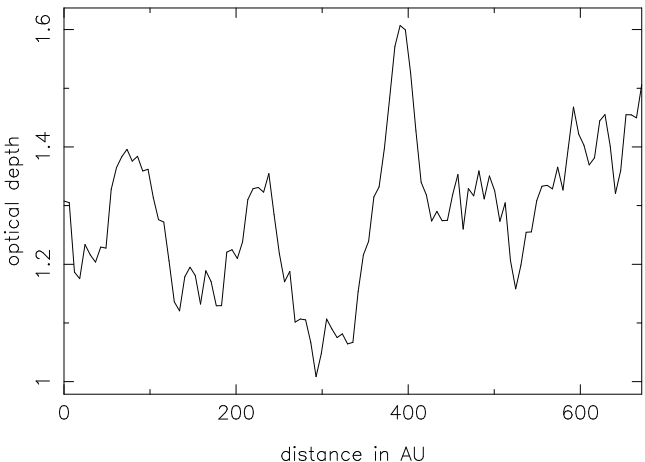,height=8.5cm,angle=-90}
\caption{ A sample section from our (1-d) simulated opacity distribution is shown.
The axis scales are calibrated consistent with the observations in the Cas-A direction
(corresponding to the scale range at longer-scales).
\label{fig:fig3}}
\end{figure}

   The important thing to note here is that the quantity $\Delta\tau_v(x_o)$, 
which relates statistically to the opacity differences observers have measured, 
has a much slower dependence on the transverse
separation ($x_o$) between the sight-lines than the {\it amplitude} at the spatial frequency ($1/x_o$)
would have in the spectrum.
Now, to make some quantitative estimates, we need
to know the details of the power spectrum. Fortunately, such details are now available from a recent
power spectrum analysis of opacity in the direction of Cas-A 
(Deshpande, Dwarakanath \& Goss 2000; hereafter DDG)  using the opacity images
measured by Bieging, Goss, \& Wilcots (1991; hereafter BGW).  The authors (DDG) report
that the power spectrum is of a power-law nature over scales ranging 
from $\sim$0.02 pc to $\sim$4 pc
and the value of $\alpha$ to be close to 2.75, significantly different
from the Kolmogorov value of 11/3. When viewed
over 0.5 km/s wide velocity channels (similar to that used in the small-scale 
structure studies being discussed here),
the rms variation in opacity 
across these images is about unity, making the structure function at 
$x_o\sim 4$ pc (corresponding to
the angular size of Cas-A and the location of the absorbing cold HI)
equal to about 2.
From this {\it calibration} and the value of $\alpha$ as suggested by the Cas-A data, 
it follows
that optical depth differences of typically 0.2 (rms) should be in fact expected at transverse
separations of $\sim$1000 AU. Given the power-law index of the structure function, at 100 AU
separation, the expected (rms) differences would drop by a factor 
of $\sim$2. To assess further
the expected opacity differences, we should examine ideally the (probability density) 
distribution of the
expected magnitudes of differences for each value of $x_o$, the structure function itself representing
the second moments of such distributions as a function of $x_o$. 
Unfortunately, the observations of BGW have 7$^{''}$ smoothing 
which limits the range of scales (at the
smaller-scale end) that we would like to examine.
Hence, simulations avoiding such smoothing were considered (see Fig. 1).
A more detailed discussion on this and related issues is given in DDG (2000).
A complex hermitian symmetric spectrum was simulated with the contributions (the real and imaginary parts)
at different spatial frequency $f_s$  
represented by uncorrelated random numbers following zero-mean
Gaussian statistics having variance matched to the $f^{-\alpha}_s$ power law. Such a spectrum over the
512$\times$512 matrix (in the 2-d case, and a $2^{18}$ point array in the 1-d case) was Fourier transformed 
to obtain a simulated $\tau$-distribution and suitably
scaled to have an rms of unity. The spatial extent of the distribution is assumed to 
correspond to
$\sim$ 4 pc, consistent with the data in Cas-A direction. Using the  simulated
distribution of opacity (similar to that observed by BGW in the Cas-A direction),
we have estimated, as a function of the spatial separation, two quantities
indicative of the related (one-sided) probability distribution of the absolute differences. 
These are the rms value (i.e.  the square-root of the structure function) and the
maximum value of the magnitudes of opacity differences.
The result of this estimation is shown in Fig. 2 and it is clear that
occasionally the opacity difference can be nearly an order of magnitude higher than
the rms values (related to the structure function).
This makes the {\it detected} differences hardly surprising and therefore, they
should be treated as only consistent with a single power-law spectral description of the opacity 
distribution (e.g. as derived in the Cas-A direction). Fig. 3 shows a small section of the
simulated distribution as an example of the expected variation on small transverse scales. 
A detailed examination of several realizations of such  distributions shows that 
the optical-depth variations of 0.2-0.4 across 
the relatively small tranverse separation (50-100 AU) as seen in Fig. 3, are not very rare,
but would be expected with $\sim$10\% probability. 

      We would like to re-emphasize here that in opacity difference
measurements with a transverse separation $x_o$, one is {\it probing} a point of 
the structure function
and {\it not} a point in the spectrum of opacity distribution. In general, in fact all scales 
in the spectrum contribute to such a measurement, except, of course, the uniform component and,
ironically, the scales $x_o$ and its integral submultiples that contribute nothing.  
In detail, the structure function value at $x_o$ is a result of a sum of the contributions at 
{\it all}
spatial frequencies in the power spectrum after modulation by $1-Cos(2\pi f_x x_o)$, and 
considering an ensemble average of such sums\footnote{More formal analytical expressions 
involving Bessel functions are commonly used. See, for example, Cordes, Weisberg \& Boriakoff (1985).}
 corresponding to all possible orientations
and locations of vectors of length $x_o$ ( along with $f_x,f_y$ axes). 
This modulation is simply a result of the two-point difference measurement. The modulating
function has its first peak at $f_x = 1/(2x_o)$ and at odd multiples of it then on.
When the power spectrum is red (i.e. $\alpha$ is positive) and steep, even the highly attenuated contributions (due to the
modulation) from the low spatial frequencies (long scales) can, and do indeed, dominate in the 
net contribution, making an equivalent scale being probed much larger than $x_o$.
One may estimate
the equivalent scale size by considering a weighted average over all scales,
where weights are determined by the associated values of the modulated power spectrum.
In a simple estimation,
for example, considering a 2-d spectrum over $|f_s|\le 1/x_o$, we find the
equivalent scale to be nearly one order of magnitude longer than $x_o$. 
A proper estimation should include the full spectrum. However, since such an equivalent scale has little
physical meaning, we have not pursued such an estimation further.
In summary, the earlier interpretations of the opacity difference observed at a 
transverse separation $x_o$ as being a result of a cloudlet of the size same 
as the separation $x_o$ appear, to us, erroneous. 

\subsection{The over-dense (\& over-pressured) HI cloudlets ?}

One of the major mysteries which owes its origin to the above mentioned misinterpretation is
that of the observed opacity changes combined with an {\it assumed}
{\it longitudinal scale} implying 
highly over-dense ($n_H \sim 10^{4}-10^{5}$ $cm^{-3}$)
and, consequently, over-pressured cloudlets. Even if the observed opacity changes were
to be accepted at their face value, their implication needs to be reinterpreted, since
a) as emphasized in the previous section, the observed variation is contributed by 
the whole range of scales and not by just one particular scale same as the transverse separation 
probed, and
b) the measured value can not be directly associated with a particular longitudinal scale.
These considerations are applicable not just to the ``two-point comparisons'' (i.e. variations
expected across a 1-d cut), but also to
the features observed in the 2-d images of the opacity.
The variations like those apparent in our 1-d simulation (Fig. 3)
would be equally probable in an equivalent
2-d image where they would appear as one-dimensional features. 
The {\it edges} or the elongated features apparent in {\it only some} of the
2-d opacity-images (Davis, Diamond \& Goss 1996; Faison et al. 1998; Faison 1999) 
are therefore not at all surprising, whereas an opacity variation feature that is narrow
in both dimensions should be considered relatively rare.

If one wants to estimate, using the data, the properties of the small-scales in the HI
distribution,
then the following is one correct way to proceed. One treats the measured value of opacity
difference or of the associated HI column-density change as just an estimate of the associated
structure function at the probed transverse separation. From this, and with some knowledge
of the spectrum (or the structure function itself), it is possible to estimate the implied
power in the same scale as the transverse separation probed. Now, this power (from the power
spectrum) or the amplitude of the {\it ripple} corresponding to that scale, 
can justifiably be interpreted in
terms of the associated {\it fluctuating} optical depth or density. More formally,
with the measurement giving an estimate of (square-root of) the structure function $S_{\tau_v}(x_o)$, 
the optical depth variation associated with
the scale $x_o$ is to be estimated as the (square-root of) power spectral contribution
$<P_{\tau_v}(f_s = \frac{1}{x_o})>$, a 
value significantly smaller than the former.  For example, using 
the relevant values observed in the Cas-A direction, a observed change of 0.2 in the optical
depth between two sight-lines with a transverse separation of about 1000 AU, would imply a
$\tau$ fluctuation on the scale of $\sim$1000 AU to be about $10^{-5}$. 
Assuming a velocity-channel width of 0.5 km/s \& a spin temperature of about 100 K,
the contribution (or deficit) in the volume-density
from that scale would be less than 0.1 $cm^{-3}$, very much smaller than what
the earlier interpretations would suggest !
This value would be even smaller when the possible  
statistical enhancement due to the finite {\it thickness} of the medium 
is accounted for. For example, if the  contribution is from
more than one, say, N layers along the sight-line, then the corresponding contribution to
volume density would be $\sqrt N$ times smaller.
It follows that
the well understood parsec scale would contribute an HI volume density 
of $\sim$ 1 $cm^{-3}$
or smaller,
in good agreement with relevant observations.
Of course, the actual density at given spatial location would be a sum total of such 
contributions also from a hierarchy of scales longer and shorter than $x_o$. 
It is easy to show that such contributions to volume density
at a given spatial point would follow a power law as a function of the scale-size. For example,
in the Cas-A direction, it may be expressible as 
$\Delta n_H(x_o) \sim A x^{\frac{\alpha-\beta}{2}}_o$, where $x_o$ is expressed in parsec,
$A$ is a constant close to unity and
$\beta$ is between 1 to 2 depending on whether our sight-line encounters contributions
many (of the order of $1/x_o$, with $x_o$ in pc) or just one layer of scale $x_o$, respectively.
Even in the worst case, i.e. when $\beta=2$, the rms fluctuations in the volume density,
estimated on the parsec scales,
would be below 100 $cm^{-3}$. The more likely value of $\beta$ in the sub-parsec regime of scales
is closer to unity, and then the volume density fluctuations would be moderate, 
with an rms of $\sim$10 $cm^{-3}$. The detailed quantitative picture may differ between different
directions in the Galaxy by a factor of 10 or less, and would hopefully be revealed by future
suitable observations \& a careful interpretation.

To conclude, there appears to be no compelling observational evidence for the
so called ``highly over-dense small scale structure'',
and even the observations probing small transverse scales 
are not at all inconsistent with what we would expect by extrapolating from the better
studied range of large and moderate spatial scales in the cold neutral medium.

\subsection{The reported measurements versus uncertainties}

    So far, we have taken the reported observations (estimations) of the opacity differences
at their face value. However, certain uncertainties inherent to the measurements are worth 
noting.
As we have discussed earlier, any observed line-visibility
in VLBI observation results from structure in both the background source and the opacity 
as seen by a given interferometer baseline. The interpretation can become more complicated
when only a limited number of baselines are used and can even be misleading if any changes in the relative
orientation of the baselines (in addition to its projected length) are not accounted for.
As was already pointed out (Radhakrishnan \& Deshpande 1990; unpublished), 
the uniqueness of the interpretation of Diamond et al. (1989) becomes debatable on these grounds.
Later similar observations however have 
resolved the possible ambiguities by actually mapping
the opacity distribution across the extra-galactic sources and hence the estimated opacity
differences can be considered reliable.

One general but important aspect, addressed earlier by Deshpande et al. (1992), is that of the
contribution from the HI emission to the measurement uncertainty (relevant to both
the interferometric and pulsar probes). Even if a VLBI baseline resolves out the large scale
HI emission, each of the elements of the interferometer does respond to the HI emission
contribution and the equivalent system temperature of the interferometer can be significantly
higher in the corresponding spectral channels compared to that for only continuum contribution.
As for the single dish observation, the HI emission contribution to the estimation uncertainty
is rather obvious. 
%More recent interferometric measurements (Davis, Diamond \& Goss 1996; Faison et al. 1998; 
The measurements in the pulsar directions (Deshpande et al. 1992; Frail et al. 1994) as well as
more recent interferometric measurements (Davis, Diamond \& Goss 1996; Faison et al. 1998; 
Faison 1999) do explicitly take into account the HI emission contribution.
Interestingly, the system temperatures in the line channels also depend, in principle, 
on the optical depths at the corresponding frequencies. This can be a significant effect 
when the background sources make dominant contribution to the system temperature (in continuum 
channels). In such cases, higher optical depths result in significantly
smaller system temperature in the corresponding spectral channels.

\section{Discussion}

In this paper we have addressed some aspects related to the observation, analysis and
interpretation of opacity differences across small (sub-parsec) transverse scales.
Although the considerations we raise are simple-minded, they appear to have serious implications
that argue against certain interpretations, such as those suggesting the so called small-scale
structure of highly over-dense cold HI cloudlets as being responsible for the observed opacity 
differences. 

We have emphasized the need for recognizing the nature of the actual 
quantity one measures through the probes that have been employed and that almost all scales contribute to
the measured opacity differences. 
But, so far, the existing studies have 
misinterpreted the observed opacity difference between two 
sight-lines of the associated transverse separation as due to opacity structure on {\it that 
scale in three-dimensions}, and therefore, it is not surprising that the implied 
volume densities appear extra-ordinarily large.
As illustrated in the earlier sections of this paper, the observations
appear consistent (both, qualitatively and quantitatively) with a single power-law description
of the HI distribution over
the entire relevant range of scales and do not imply any mysterious structure.
In our simple picture, the spectral behaviour 
studied up to moderate scales (e.g. $\sim$ 0.1 pc scale as by DDG, 2000) 
is assumed to extend with the same
power-law index to the 10 AU scale. We are aware of the study by Croviser, Dickey \&
Kazes (1985) that claimed a cutoff in structures below 0.2 pc. Considering 1) their 
method in which signatures of small-scale structure were expected to show up close 
to the zero-velocity,
2) the major absorption line features (differences in which were probed) were well away from
zero-velocity, and
3) the velocity resolution was coarse,
we think that their study suffers from serious selection against structures 
smaller than about 0.5 pc.

The recent suggestion by Heiles (1997) did for the first time distinguish between
the transverse scale probed and the longitudinal scale for estimating the volume density,
but by invoking ``thin''  structures such as filaments \& sheets that  should
preferentially align along sight-lines. As we have discussed, the distinction between
the transverse scale probed and the {\it corresponding equivalent} longitudinal scale
appears to be rather simple and more inherent to the basic measurement than any geometrical shapes
would imply.
While any anisotropy in structures is likely to increase (statistically) the expected 
magnitude of the opacity differences (and elongation factors up to $\sim$2 may be common)
it does not appear necessary for understanding the available observations. 
In any case, the structures 
suggested by  Heiles may be difficult to produce and maintain, particularly in the certain
alignment that they need to have with {\it our} sight-lines.

The velocity spread associated with turbulence can produce additional corrugations in
the opacity distribution when viewed over velocity channels narrow compared to the
spread due to turbulence (as is normally the case). It would be instructive to assess
such an effect quantitatively.

It may be important to note that the radio/optical observations of opacity changes in 
other species
({\it e.g.,} a. H$_2$CO and OH by Moore \& Marscher 1995; 
             b. NaI and CaII by Mayer \& Blades 1996, and Watson \& Meyer 1996)
should be interpreted 
much the same way as we have discussed in the
context of HI and then a similar puzzle these observations appeared to have raised should stand 
resolved. The discussion in this paper is relevant also to the dispersion measure changes detected
in pulsar directions and in general to any situation involving a similar probe.

Although in the course of our discussion we have drawn upon the HI data in the Cas-A direction
as a useful example, we do recognize the possibility that the atomic medium in other
directions may have quite different column densities as well as power-spectral
descriptions from that in the Cas-A direction. 
However, the main issues we have addressed are of a more general nature 
and are not crucially based on the quantitative estimates from the Cas-A data.
The HI emission line study by Green (1993) indicates that the power spectra 
(and structure functions) may be less steep in some directions than that for
cold HI in Cas-A direction. A recent study of HI in the Small Magellanic Cloud
suggests a relatively steeper power spectrum (Stanimirovic {\it et al.} 1999). 
If the power spectra of scale distribution in cold
HI also have a similar variation, then in some directions we should expect even
larger opacity differences on small transverse separations (much more than even
those seen in recent VLBI studies). For example, a change of 0.1 in $\alpha$ would
increase the {\it expected} opacity changes on the AU scales by a factor of $\sim$2. 

Further investigations should benefit from using the
 available data for a systematic estimation of the structure function associated
with the opacity distribution over the relevant range of transverse scales.
HI absorption measurements on moderate size, bright background sources should help
extending the direct power spectral analysis (e.g in the case of Cas-A) to
intermediate and small transverse scales. 

\subsection*{Acknowledgements:}
The author thanks K. R. Anantharamaiah, Barry Clark, Vivek Dhawan, 
K. S. Dwarakanath, Dale Frail, Miller Goss, Carl Gwinn, 
Rajaram Nityananda and V. Radhakrishnan for interesting discussions \& 
many useful comments on the manuscript, and Peter Scheuer for his helpful suggestions
on the introductory passages.
The author gratefully acknowledges also the inspiration, education and encouragement
received from V. Radhakrishnan through numerous stimulating discussions 
on the issues addressed in the paper.

\bigskip

\centerline {\bf References}

\parhang Bieging, J. H., Goss, W. M., \& Wilcots, E. M. 1991, {\it ApJSS}, {\bf 75}, 999. \vskip 1pt

\parhang Bregman, J. D., Troland, T. H., Foster, J. R., Schwarz, U. J., Goss, W. M., \& Heiles, C. 
1983, {\it A\&A}, {\bf 118}, 157.  \vskip 1pt 

\parhang Clark, B. G. 1965, {\it ApJ}, {\bf 142}, 1398.  \vskip 1pt 

\parhang Clark, B. G., Radhakrishnan, V. \& Wilson, R. W. 1962, {\it ApJ}, {\bf 135}, 151.  \vskip 1pt 

\parhang Cordes, J. M., Weisberg, J. M., \& Boriakoff, V. 1985, {\it ApJ}, {\bf 288}, 221.  \vskip 1pt 

\parhang Croviser, J., Dickey, J. M., \& Kazes, A. 1985, {\it A\&A}, {\bf 146}, 223.  \vskip 1pt 

\parhang Davis, R. J., Diamond, P. J., \& Goss W. M. 1996, {\it MNRAS}, {\bf 283}, 1105. \vskip 1pt

\parhang Deshpande, A. A., Dwarakanath, K. S. \& Goss, W. M. 2000, {\it ApJ}, preprint.

\parhang Deshpande, A. A., McCulloch, P. M., Radhakrishnan, V., \& Anantharamaiah, K.R.
1992, {\it MNRAS}, {\bf 258}, 19p. \vskip 1pt 

\parhang Diamond, P. J., Goss, W. M., Romney, J. D., Booth, R. S., Kalbera, P. M. W., \&
Mebold, U. 1989, {\it ApJ}, {\bf 347}, 302. \vskip 1pt 

\parhang Dickey, J. M., \& Lockman, F. J. 1990, {\it ARA\&A}, {\bf 28}, 215.  \vskip 1pt 

\parhang Dieter, N. H., Welch, W. J., \& Romney J. D. 1976, {\it ApJ}, {\bf 206}, L113.  \vskip 1pt 

\parhang Faison, M. D. 1999, Ph.D. thesis, Univ. Wisconsin-Madison. \vskip 1pt

\parhang Faison, M. D., Goss, W. M., Diamond, P. J. \& Taylor, G. B. 1998, {\it AJ}, 
{\bf 116}, 2916.\vskip 1pt

\parhang Frail, D. A., Cordes, J. M., Hankins, T. H. \&  Weisberg, J. M. 1991, {\it ApJ}, {\bf 382}, 168. 
\vskip 1pt 

\parhang Frail, D. A., Weisberg, J. M., Cordes, J. M., \& Mathers, C. 1994, {\it ApJ}, {\bf 436}, 144. 
\vskip 1pt 

\parhang Green, D. A. 1993, {\it MNRAS}, {\bf 262}, 327.  \vskip 1pt 

\parhang Greisen, E. W. 1973, {\it ApJ}, {\bf 184}, 363.  \vskip 1pt 

\parhang Heiles, C. 1997, {\it ApJ}, {\bf 481}, 193.  \vskip 1pt 

\parhang Lazarian, A. 1995, {\it A\&A}, {\bf 293}, 507.  \vskip 1pt 

\parhang Lee, L. C., \& Jokippii, J. R. 1975, {\it ApJ}, {\bf 196}, 695.  \vskip 1pt 

\parhang Mayer, D. M, \& Blades, J. C. 1996, {\it ApJ}, {\bf 464}, L179.  \vskip 1pt 

\parhang Moore, E. M., \& Marscher, A. P. 1995, {\it ApJ}, {\bf 452}, 671. \vskip 1pt 

\parhang Schwarz, U. J., Troland, T. H., Albinson, J. S., Bregman, J. D., Goss, W. M., \& Heiles, C.
1986, {\it ApJ}, {\bf 301}, 320.  \vskip 1pt 

\parhang Stanimirovic, S., Staveley-Smith, L., Dickey, J. M., Sault, R. J.,
\& Snowden, S. L. 1999, {\it MNRAS}, {\bf 302}, 417. \vskip 1pt

\parhang Watson, J. K., \& Meyer, D. M. 1996, {\it BAAS}, {\bf 28}, 892.  \vskip 1pt

\end{document}